# Physical-type correctness in scientific Python

Marcus Foster, Retired (eResearch support)

Sean Tregeagle, Software Developer

## Abstract

The representation of units and dimensions in informatics systems is barely codified and often ignored. For instance, the major languages used in scientific computing (Fortran, C and Python), have no type for dimension or unit, and so physical quantities are represented in a program by variables of type real, resulting in the possibility of unit or dimensional errors. In view of this danger, many authors have proposed language schemes for unit-checking and conversion. However, since many physical quantities have the same units, it is possible for a block of code to be unit-compatible, but still physically meaningless. We demonstrate the limitations of three Python unit-libraries and present a justification and method for checking kind-of-quantity.

## Recapitulation

Quantity equations such as $pV = nRT$, $F = G\, m_1 m_2/r^2$ and $E = mc^2$ express physical phenomena and are the basis of most scientific calculations and models. Quantities do not exist in isolation but are part of a quantity system comprising base and derived quantities. A base quantity was formerly called a dimension; now 'dimension' expresses the combination of base quantities in a derived quantity, for example in a mass-length-time system, the dimensions of energy are mass*length$^2$/time$^2$ ($\mathbf{ML^2T^{-2}}$).

When assigning or algebraically manipulating quantities, the two basic rules of compatibility are:
1. Quantities of different dimensions or units cannot be added (or subtracted).
2. The dimensions and units of the LHS of a quantity assignment must equal the dimensions and units of the RHS.

Since there is a many-to-one correspondence from unit-of-measure to dimension, satisfying unit-compatibility, also satisfies dimensional-compatibility. So, let's call these two rules of compatibility, Type 1 and Type 2 unit compatibility. Of course, if we wish to algebraically manipulate different units, we need to track both their dimensions and scaling.

## Introduction

The representation of units and dimensions in informatics systems is barely codified and often ignored [1]. For instance, the major languages used in scientific computing (Fortran, C and Python), have no type for dimension or unit, and so physical quantities are represented in a program by variables of type real. This immediately raises the possibility of dimensional errors, for instance where we inadvertently add a force variable **F_centripetal** to an energy variable **E_thermal**, or assign an expression of dimension length/time to a variable of dimension length. Similarly, the dimensions of an expression may be compatible, but a unit error occurs when we add an energy variable **E_kinetic** in kJ to another energy variable **E_potential** in ft-lbf.

Automatic checking and conversion of physical types increases program safety (possibly at the expense of simplicity or speed). A proposal for incorporating unit-checking in Fortran compilers [2], states that the "third most detectable type of error in scientific software is incorrect use of physical units of measurement". Other schemes for manipulating and checking units and dimensions include source-code annotation [3], function-libraries [4] and class-libraries [5] [6].

# Unit-checking in Python

Let's demonstrate three Python units packages.

Pint [4] is "a Python package to define, operate and manipulate physical quantities: the product of a numerical value and a unit of measurement. It allows arithmetic operations between them and conversions from and to different units".

In the program *Pint1*, Pint converts units/prefixes in line 7 and detects incompatible units in line 8.

```
1 # Pint1
2 import pint
3 u=pint.UnitRegistry()
4 distance1 = 10.5*u.cm
5 distance2 = 3.3*u.ft
6 speed = 42.0*u.km/u.hour
7 print(distance1 + distance2)
8 print(distance1 + speed)
9 # end Pint1
```

The (simplified) Pint output is:
```
111.084 centimeter
  File "D:\KOQ\Pint1.py", line 8, in <module>
    print(distance1 + speed)
DimensionalityError: Cannot convert from 'centimeter' ([length]) to
'kilometer / hour' ([length] / [time])
```

ScientificPython [5] is a large library of scientific routines, and contains a class-module PhysicalQuantities, which similarly converts units/prefixes in line 6 and detects incompatible units in line 7 of program *PQ1*.

```
1 # PQ1
2 from Scientific.Physics.PhysicalQuantities import PhysicalQuantity as p
3 distance1 = p(10.5, 'cm')
4 distance2 = p(3.3, 'ft')
5 speed = (42.0, 'km/h')
6 print(distance1 + distance2)
7 print(distance1 + speed)
8 # end PQ1
```

The (simplified) PhysicalQuantities output is:
```
111.084 cm
  File "D:\KOQ\PQ1.py", line 7, in <module>
    print(distance1 + speed)
TypeError: Incompatible types
```

Unyt [6] is a recent sophisticated and high-performance units package [7], which leverages the SymPy class library to yield an elegant and readable unit syntax. Program *unyt1* demonstrates unit conversion (line 6) and detects incompatible units (line 7).

```
1 # unyt1.py
2 from unyt import cm, ft, km, hr
3 distance1 = 10.5*cm
4 distance2 = 3.3*ft
5 speed = 42.0*km/hr
6 print(distance1 + distance2)
7 print(distance1 + speed)
8 # end unyt1.py
```

The (simplified) unyt output is:
```
111.08399999999999 cm
  File "D:\KOQ\unyt1.py", line 7, in <module>
    print(distance1 + speed)
UnitOperationError: The <ufunc 'add'> operator for unyt_arrays with units
"cm" (dimensions "(length)") and "km/hr" (dimensions "(length)/(time)") is
not well defined.
```

Note that Python is a dynamically-typed language, and so the Python interpreter will silently change the type of a variable. Thus, a Type 2 unit incompatibility cannot be detected by the interpreter, regardless of whether we use a unit-checking module. Program *Pint2* demonstrates this below:

```
1 # Pint2
2 import pint
3 u=pint.UnitRegistry()
4 distance1 = 10.5*u.cm
5 distance2 = 3.3*u.ft
6 speed = 42.0*u.km/u.hour
7 speed = distance1 + distance2
8 print(speed)
9 # end program Pint1
```

The Python output is:
```
111.084 cm
```

## The shortcomings of unit-checking

All the schemes described above, use a simplified representation of physical types, and do not incorporate 'kind-of-quantity' (KOQ), which is another attribute [8] of a physical type, at a higher abstraction than dimension. Thus, two quantities may have the same dimensions, but can be different kinds of quantities. Familiar examples are energy and torque, both of dimension **$ML^2T^{-2}$**, and heat capacity and entropy, both of dimension **$ML^2T^{-2}\Theta^{-1}$**. The number of kinds-of-quantities is unlimited; for example, a recent handbook [9] catalogues 1200 different kinds of 'characteristic numbers' (also known as dimensionless parameters) of dimension '1', and there are an unlimited number of relative quantities (e.g. refractive index, mass fraction, friction factor, Mach number) also of dimension '1'.

The compatibility rules for dimensions and units, also apply to kind-of-quantity: it is meaningless to add a variable of type torque to a variable of type energy (let's call this a Type1 KOQ error), or assign an expression of type torque to a variable of type energy (a Type 2 KOQ error). A simple example of a Type 2 KOQ error is the incorrect analysis of a turbine, of moment-of-inertia $I$ (SI unit: kg·m$^2$), rotating with an angular velocity of $\omega_1$ (s$^{-1}$) with a torque $T$ (kg·m$^2$·s$^{-2}$) applied for duration $t$ (s). The initial kinetic energy $E_1$ is defined as $E_1 = 0.5 * I * \omega_1^2$. It is easy to code this quantity equation incorrectly as $E_1 = 0.5 * I / t^2$, where the units of both sides of the assignment (kg·m$^2$·s$^{-2}$) are compatible, but the kind-of-quantity on the RHS is undefined.

So while the Pint, PhysicalQuantities and unyt packages are a great convenience and safety net for the scientific programmer, because they track only dimensions and units, they cannot necessarily detect kind-of-quantity errors.

Consider the following simplified real-world problem. A 70 MW (hydraulic power) hydro-generator has a moment of inertia of 16000 kg.m$^2$. The turbine is accelerated at no load from idle of 10 rev/min to synchronous speed of 93.75 rev/min over 3 minutes. Find the average torque applied to the generator shaft. Find the final kinetic energy of the generator and final torque applied, at full electrical load.

Any unit check of the program *KOQErrors-Undetected* below, cannot detect the Type 1 KOQ error in line 13, nor the Type 2 KOQ error in line 15. Pint, for example, does not return any error messages for this program.

```
1 # KOQErrors-Undetected.py
2 import math
```

```
 3 import pint
 4 u=pint.UnitRegistry()
 5 power=70.e6*u.kg*u.meter**2/u.second**3
 6 duration=180.*u.second
 7 I=16000.*u.kg*u.meter**2
 8 av1=10./60.*2.*math.pi/u.second
 9 av2=93.75/60.*2.*math.pi/u.second
10 energy1=0.5*I*av1*av1
11 torque_avg=(av2-av1)*I/duration
12 # Pint cannot detect this Type 1 KOQ error: meaningless quantity
addition:
13 energy2=energy1+torque_avg
14 # Pint cannot detect this Type 2 KOQ error: meaningless quantity
multiplication:
15 energy2=0.5*I/(duration*duration)
16 # correct final energy is given by
17 energy2=0.5*I*av2*av2
18 # final torque is given by
19 torque2=power/av2
20 # end KOQErrors-Undetected.py
```

## Detecting KOQ errors in Python

Detecting a Type 1 KOQ error requires a straightforward compatibility check, given a declaration of each variable's KOQ. But how do we detect a Type 2 KOQ error? Since there is a many-to-one correspondence from KOQ to dimension, a quantity's KOQ cannot be inferred from its dimensions. KOQ represents the axiomatic relationships in a quantity equation; thus, these relationships also need to be declared in a program.

Quant [10] is a proof-of-concept Python package for detecting KOQ errors in Python code. Expressions are 'annotated' with their KOQ names, allowing Quant to flag Type 1 KOQ errors. It achieves this by maintaining a simple dictionary of names that it uses to check the operands of an addition or subtraction. In addition, Quant also allows KOQ relations to be declared which it uses to detect Type 2 KOQ errors. Detecting Type 2 errors is more complex, and requires both a dictionary of relations, and run-time interception of mathematical operations to record both the operators and their operands. On completion, Quant compares the expression executed with the dictionary of relations and flags any mismatch (i.e., a Type 2 KOQ error). Quant's expression comparison is currently simplistic and can be fooled. A more robust implementation would need to be able to simplify and/or re-arrange expressions to recognise equivalence in more complex cases. Similarly, Quant only supports basic arithmetic operators at this time, but it could be extended to cater for additional operators and functions.

The Quant syntax for declaring KOQ relationships and KOQ names is:

```
quant.KOQRegistry().KOQRelation("KOQ1", "KOQ2 [*|/] KOQ3 [*|/] KOQ4...")
<variable>=quant.KOQRegistry().KOQ("KOQ1", <expression>)
```

Quant is demonstrated below, using the same base code as the example above. It tracks the KOQ attribute of each physical-type variable against the KOQ relations declared for this system. Since this physical system has two modes (accelerated and constant angular velocity), the KOQ relation for torque is redefined toward the end of the program

```
1 # KOQErrors-Detected.py
2 import math
3 import pint
4 u=pint.UnitRegistry()
5 import quant
```

```
 6 q=quant.KOQRegistry()
 7 q.KOQRelation("TORQUE","AV*MOI/TIME")
 8 q.KOQRelation("ROTENERGY","MOI*AV*AV")
 9 power=q.KOQ("POWER",70.e6*u.kg*u.meter**2/u.second**3)
10 duration=q.KOQ("TIME", 180.*u.second)
11 I=q.KOQ("MOI",16000.*u.kg*u.meter**2)
12 av1=q.KOQ("AV",10./60.*2.*math.pi/u.second)
13 av2=q.KOQ("AV",93.75/60.*2.*math.pi/u.second)
14 energy1=q.KOQ("ROTENERGY",0.5*I*av1*av1)
15 torque_avg=q.KOQ("TORQUE",(av2-av1)*I/duration)
16 # Quant detects this Type 1 KOQ error: meaningless quantity addition:
17 energy2=energy1+torque_avg
18 # Quant detects this Type 2 KOQ error: meaningless quantity multiplication:
19 energy2=q.KOQ("ROTENERGY",0.5*I/(duration*duration))
20 # correct final energy is given by
21 energy2=q.KOQ("ROTENERGY",0.5*I*av2*av2)
22 #redefine KOQRelation for constant angular velocity
23 q.KOQRelation("TORQUE", "POWER/AV")
24 # final torque is given by
25 torque2=q.KOQ("TORQUE", power/av2)
26 # end KOQErrors-Detected.py
```

Quant detects the KOQ errors in the expressions that Pint (nor PhysicalQuantities or unyt) cannot, i.e.

**<Line 17>** `TypeError: Type 1 Kind of Quantity error: ROTENERGY vs 'TORQUE'`
**<Line 19>** `TypeError: Type 2 Kind of Quantity error: 'ROTENERGY = ['MOI*AV*AV']'`

## Conclusion

Ultimately, the ongoing safety of a scientific program can be maximised by a layered approach [11], including using all available language features to trap errors. The Python language is dynamically typed, and so does not provide protection against changing the (physical) type of a variable. While this flexibility has its programming benefits, a complex scientific Python program ought to be additionally checked for physical-type correctness (i.e., unit, dimension and kind-of-quantity compatibility). There is clearly an additional programming burden and a performance penalty for performing these additional checks, so that this choice is a cost-benefit judgement by the developer.

## Caution

One last caution: incorporating KOQ-checking into software, does not make physical-type checking complete. Most unit-aware software is based on the International System of Units (SI) [12], and the SI is not a formally consistent system. The definition of the hertz is known for introducing errors of $2\pi$, and it has been proposed [13] that the hertz is not a coherent SI unit. There is ambiguity in the SI's treatment of the unit '1', which has dual meanings as a base unit for count and as a derived unit for the ratio of quantities of the same kind (and, of course, the characteristic numbers already mentioned). The SI designates plane and solid 'angle' as dimensionless quantities, although there are arguments [14] that angle has its own dimension and that what is called 'angle' is actually the *number* of radians in the (dimensioned) angle. If angle were dimensioned, for instance, then a unit-checker could detect the incorrect addition of an energy variable (N m) to a torque variable (N m/rad). Currently, calculation errors are possible when manipulating dimensionless quantities and the units radian, steradian, hertz and mole. Numerous authors have criticised these SI units, and a recent paper [15] has proposed an architectural change to the SI to resolve the ambiguity of the unit '1'.

**Marcus P. Foster,** now retired, was with Information Management and Technology, CSIRO, Australia. He maintains an interest in the (mis)representation of quantities and units in computer systems.

**Sean Tregeagle** is a Java developer, currently working for an Australian Government department.